\documentclass[9pt]{article}\usepackage{spconf,amsmath,graphicx}     

\usepackage{amsfonts}
\usepackage{color}
\usepackage{graphicx}
\usepackage[dvips]{epsfig}
\usepackage{graphics} 
\usepackage{times} 
\usepackage{amssymb}  
\usepackage{cite}
\usepackage{multirow}
\usepackage[tight,footnotesize]{subfigure}
\usepackage{algorithm}

\def\norm #1{\left\|#1\right\|}

\def\twon #1{\left\|#1\right\|_2}

\def\frobn #1{\left\|#1\right\|_{\text{F}}}

\def\atomn #1{\left\|#1\right\|_{\cA}}

\def\abs #1{\left|#1\right|}

\def\st{\text{subject to }}

\def\bC{\mathbb{C}}

\def\bS{\mathbb{S}}

\def\m #1{\boldsymbol{#1}}

\def\cA{\mathcal{A}}

\def\cD{\mathcal{D}}

\def\cK{\mathcal{K}}

\def\bee{\begin{equation}}
\def\ene{\end{equation}}

\def\beq{\begin{eqnarray}}
\def\enq{\end{eqnarray}}

\newtheorem{thm}{Theorem}

\newtheorem{defi}{Definition}

\def\equ #1{\begin{equation}#1\end{equation}}

\def\sbra #1{\left(#1\right)}
\def\mbra #1{\left[#1\right]}
\def\lbra #1{\left\{#1\right\}}

\def\tr #1{\text{tr}#1}

\def\rank #1{\text{rank}#1}
\def\spark #1{\text{spark}#1}
\def\st {\text{ subject to }}

\title{Continuous Compressed Sensing With a Single or Multiple Measurement Vectors}
\name{Zai Yang and Lihua Xie, Fellow, IEEE}
\address{School of Electrical and Electronic Engineering, Nanyang Technological University, 639798, Singapore\\
\{yangzai, elhxie\}@ntu.edu.sg \\
In \emph{IEEE Workshop on Statistical Signal Processing (SSP)}, pp. 308--311, June 2014}

\begin{document}
\ninept
\maketitle


\begin{abstract} We consider the problem of recovering a single or multiple frequency-sparse signals, which share the same frequency components, from a subset of regularly spaced samples. The problem is referred to as continuous compressed sensing (CCS) in which the frequencies can take any values in the normalized domain $\left[0,1\right)$. In this paper, a link between CCS and low rank matrix completion (LRMC) is established based on an $\ell_0$-pseudo-norm-like formulation, and theoretical guarantees for exact recovery are analyzed. Practically efficient algorithms are proposed based on the link and convex and nonconvex relaxations, and validated via numerical simulations.
\end{abstract}


\section{Introduction}
Compressed sensing (CS) studies sparse signal recovery from far fewer measurements and has brought significant impact on signal processing and information theory in the past decade. Since its development thus far has been focused on signals that can be sparsely represented under a finite discrete dictionary, limitations are present in applications such as array processing, radar and sonar, where the dictionary is typically specified by one or more continuous parameters.
In this paper, we consider the problem of recovering a sinusoidal/frequency-sparse signal, which is a superposition of a few complex sinusoids, from a subset of regularly spaced samples. The problem is referred to as continuous CS as suggested in \cite{tang2012compressed} in the sense that the frequencies of the sinusoids can take any continuous values. A systematic, convex approach is introduced in \cite{tang2012compressed} which works directly in the continuous domain and completely eliminates parameter discretization/gridding of conventional CS methods that causes basis mismatches. In the paper, it is shown that a frequency-sparse signal can be exactly recovered in the noiseless case from far fewer samples provided that the frequencies are appropriately separate. Practical solutions are then provided in \cite{yang2014gridless} in the noisy case. An important, related application is direction of arrival (DOA) estimation \cite{krim1996two}, in which improved performance is typically obtained by receiving multiple frequency-sparse signals in a time interval which share the same frequency components. Note that the method in \cite{tang2012compressed} specified for a single measurement vector (SMV) cannot process the multiple measurement vectors (MMVs) at a single step while the joint processing exploiting the so-called joint sparsity can usually improve the performance \cite{eldar2010average}. Before this paper, the only known continuous/gridless sparse method for MMVs was presented in \cite{yang2014discretization} in the context of DOA estimation based on statistical inference.

In this paper, we study the SMV and MMV continuous CS problems in a unified framework. Based on an $\ell_0$-norm-like formulation we establish a link between continuous CS and a well-studied area of low rank matrix completion (LRMC) \cite{candes2009exact} and provide a sufficient condition for exact recovery. We propose convex optimization methods for signal recovery based on the link and convex and nonconvex relaxations and present computationally efficient algorithms using alternating direction method of multipliers (ADMM) \cite{boyd2011distributed}. Numerical simulations are provided to study their phase transition phenomena and validate their usefulness in DOA estimation.

\section{Signal Recovery via Atomic $\ell_0$ Norm Minimization}

\subsection{Atomic $\ell_0$ Norm Minimization}
Suppose that we observe a number of $L$ sinusoidal signals
\equ{y_{jt}^o=\sum_{k=1}^K s_{kt}e^{i2\pi (j-1) f_k}, \quad \sbra{j,t}\in \mbra{N}\times\mbra{L}, \label{formu:observmodel1}}
denoted by matrix $\m{Y}^o=\mbra{y_{jt}^o}\in\bC^{N\times L}$, on the index set $\m{\Omega}\times\mbra{L}$, where $\m{\Omega} \subset \mbra{N}\triangleq\lbra{1,\cdots,N}$ with $M\triangleq\abs{\m{\Omega}}\leq N$ denoting the sample size of each sinusoidal signal.
Here $(j,t)$ indexes the $j$th entry of the $t$th measurement vector (or snapshot data), $i=\sqrt{-1}$, $f_k\in\left[0,1\right)$ denotes the $k$th normalized frequency, and $s_{kt}\in\bC$ is the (complex) amplitude of the $k$th component at snapshot $t$. The SMV case where $L=1$ corresponds to line spectral estimation in spectral analysis and the MMV case is common in array processing. In this paper, each column of $\m{Y}^o$ is called a frequency-sparse signal since the number of sinusoids $K$ is typically small. We are interested in the recovery of $\m{Y}^o$ (as well as the parameters $\m{f}$ and $\m{s}$ if possible) under the sparse prior given its partial or compressive measurements on $\m{\Omega}\times\mbra{L}$, denoted by $\m{Y}_{\m{\Omega}}^o$. This problem is called the continuous CS problem to distinguish with the common discrete frequency setting as suggested in \cite{tang2012compressed}. We mainly consider the noiseless case. The general noisy case will be deferred to Subsection \ref{sec:noisycase}.

We exploit sparsity to solve the ill-posed problem of recovering $\m{Y}^o$ from $\m{Y}_{\m{\Omega}}^o$. Following the literature of CS, we seek for the maximally sparse candidate for its recovery. To state it formally, we denote $\m{a}\sbra{f}=\mbra{1,e^{i2\pi f},\cdots,e^{i2\pi\sbra{N-1}f}}^T\in\bC^{N}$ and $\m{s}_k=\mbra{s_{k1},\cdots,s_{kL}}\in\bC^{1\times L}$. Then (\ref{formu:observmodel1}) can be written as
\equ{\m{Y}^o=\sum_{k=1}^K \m{a}\sbra{f_k}\m{s}_k=\sum_{k=1}^K c_k\m{a}\sbra{f_k}\m{\phi}_k, \label{formu:atomdecomp}}
where $c_k=\twon{\m{s}_k}>0$ and $\m{\phi}_k=c_k^{-1}\m{s}_k$ with $\twon{\m{\phi}_k}=1$. Let $\bS^{2L-1}=\lbra{\m{\phi}: \m{\phi}\in\bC^{1\times L}, \twon{\m{\phi}}=1}$ denote the unit $2L-1$-sphere. We define the continuous dictionary or the set of atoms
\equ{\cA\triangleq \lbra{\m{a}\sbra{f,\m{\phi}}=\m{a}\sbra{f}\m{\phi}: f\in\left[0,1\right), \m{\phi}\in\bS^{2L-1}}.\label{formu:atomset}}
It is clear that $\m{Y}^o$ is a linear combination of a number of atoms in $\cA$. We define the atomic $\ell_0$ (pseudo-)norm of some $\m{Y}\in\bC^{N\times L}$ as the smallest number of atoms that can express it:
\equ{\norm{\m{Y}}_{\cA,0}
=\inf\lbra{\cK: \m{Y}=\sum_{k=1}^{\cK} c_k\m{a}_k, \m{a}_k\in\cA, c_k>0}. \label{formu:AL0}}
So we propose the following problem for signal recovery:
\equ{\min_{\m{Y}} \norm{\m{Y}}_{\cA,0}, \st \m{Y}_{\m{\Omega}}=\m{Y}^o_{\m{\Omega}}, \label{formu:AL0min}}
where $\m{Y}_{\m{\Omega}}$ takes the rows of $\m{Y}$ indexed by $\m{\Omega}$.

\subsection{Spark of Continuous Dictionary}
To analyze the atomic $\ell_0$ norm minimization problem in (\ref{formu:AL0min}), we generalize the concept of spark to the case of continuous dictionary. We define the following continuous dictionary with respect to the index set $\m{\Omega}$:
$\cA_{\m{\Omega}}^1\triangleq \lbra{\m{a}_{\m{\Omega}}\sbra{f}: f\in\left[0,1\right)}$.

\begin{defi}[Spark of continuous dictionary] Given the continuous dictionary $\cA_{\m{\Omega}}^1$, the quantity spark of $\cA_{\m{\Omega}}^1$, denoted by $\spark\sbra{\cA_{\m{\Omega}}^1}$, is the smallest number of atoms of $\cA_{\m{\Omega}}^1$ which are linearly dependent.
\end{defi}

\begin{thm} We have the following results about $\spark\sbra{\cA_{\m{\Omega}}^1}$:
\begin{enumerate}
 \item $2\leq\spark\sbra{\cA_{\m{\Omega}}^1}\leq M+1$,
 \item $\spark\sbra{\cA_{\m{\Omega}}^1}= 2$ if and only if the elements of
  \equ{\m{\cD}\triangleq\lbra{m_1-m_2: m_1,m_2\in\m{\Omega},m_1\geq m_2}}
      is not coprime, and
 \item $\spark\sbra{\cA_{\m{\Omega}}^1}= M+1$ if $\m{\Omega}$ consists of $M$ consecutive integers.
\end{enumerate}
 \label{thm:fullspark}
\end{thm}

Theorem \ref{thm:fullspark} presents the range of $\spark\sbra{\cA_{\m{\Omega}}^1}$ with respect to the sampling index set $\m{\Omega}$. Readers are referred to \cite{yang2014exact} for its proof and those of the rest results due to the page limit. A sufficient and necessary condition is provided under which $\spark\sbra{\cA_{\m{\Omega}}^1}$ achieves the lower bound 2. Note, however, that such $\m{\Omega}$ is rare. For example, when $\m{\Omega}$ is selected uniformly at random, the probability that the condition holds is 0 whenever $M>\frac{N}{2}$. It is less than $1.2\times10^{-3},\, 1.8\times10^{-7},\, 3.2\times10^{-12}$ when $N=100$ and $M=10,\,20,\,30$ respectively. A sufficient (but unnecessary) condition is also provided under which $\cA_{\m{\Omega}}^1$ achieves the upper bound $M+1$.

\subsection{Sufficient Condition for Exact Recovery} \label{sec:l0exactrecovery}

We provide theoretical guarantees of the atomic $\ell_0$ norm minimization in (\ref{formu:AL0min}) for frequency recovery in this subsection. In particular, we have the following result, which can be considered as a continuous version of \cite[Theorem 2.4]{chen2006theoretical}.

\begin{thm} $\m{Y}^o=\sum_{k=1}^K c_k\m{a}\sbra{f_k,\m{\phi}_k}$ is the unique optimizer to (\ref{formu:AL0min}) if
\equ{K< \frac{\spark\sbra{\cA_{\m{\Omega}}^1}-1+\rank \sbra{\m{Y}_{\m{\Omega}}^o}}{2}. \label{Kbound}}
Moreover, the atomic decomposition above is the unique one satisfying that $K=\norm{\m{Y}^o}_{\cA,0}$.
\label{thm:AL0_guanrantee}
\end{thm}

Theorem \ref{thm:AL0_guanrantee} shows that the proposed atomic $\ell_0$ minimization problem can recover a frequency-sparse signal with sparsity $K< \frac{1}{2}\spark\sbra{\cA_{\m{\Omega}}^1}$ in the SMV case. As we take more measurement vectors, we have a chance to recover more complex signals by increasing $\rank \sbra{\m{Y}_{\m{\Omega}}^o}$, which is practically relevant in array processing applications. In fact, the sparsity $K$ can be as large as $M-1$ with an approximate choice of $\m{\Omega}$.

\subsection{Finite Dimensional Characterization via Rank Minimization}
The optimization problem in (\ref{formu:AL0min}) is computationally infeasible given the infinite dimensional formulation of the atomic $\ell_0$ norm in (\ref{formu:AL0}). We provide a finite dimensional formulation in the following result.

\begin{thm}
$\norm{\m{Y}}_{\cA,0}$ defined in (\ref{formu:AL0}) equals the optimal value of the following rank minimization problem:
\equ{\min_{\m{W}\in\bC^{L\times L},\m{u}\in\bC^{N},\m{U}\geq\m{0}} \rank\sbra{\m{U}}, \st \m{U}=\begin{bmatrix}\m{W} & \m{Y}^H \\ \m{Y} & T\sbra{\m{u}}\end{bmatrix}, \label{formu:AL0_rankmin}}
where $T\sbra{\m{u}}$ denotes a (Hermitian) Toeplitz matrix with its first row specified by $\m{u}^T$, and $\m{U}\geq\m{0}$ means that $\m{U}$ is positive semidefinite.
\label{thm:AL0_rankmin}
\end{thm}


Theorem \ref{thm:AL0_rankmin} presents a rank minimization problem to characterize the atomic $\ell_0$ norm. It follows that (\ref{formu:AL0min}) is equivalent to the following LRMC problem:
\equ{\begin{split}
&\min_{\m{Y},\m{W},\m{u},\m{U}\geq\m{0}} \rank\sbra{\m{U}},\\
&\st \m{U}=\begin{bmatrix}\m{W} & \m{Y}^H \\ \m{Y} & T\sbra{\m{u}}\end{bmatrix}, \m{Y}_{\m{\Omega}}=\m{Y}^o_{\m{\Omega}},\end{split} \label{formu:rankmin_CCS}}
where we need to recover a (structured positive semidefinite) low rank matrix $\m{U}$ with partial access to its entries. As a result, we establish a link between continuous CS and the well studied area of LRMC, which enables us to study the continuous CS problem by borrowing ideas in LRMC. Note that a similar rank minimization problem is presented in \cite{tang2012compressed} in the SMV case, where the rank is put on the matrix $T\sbra{\m{u}}$ rather than the full matrix $\m{U}$ in (\ref{formu:rankmin_CCS}). This difference obscures the link between continuous CS and LRMC which, we will see later, plays an important role in this paper.

\section{Signal Recovery via Relaxations}
\subsection{Convex Relaxations} \label{sec:convexrelax}

The atomic $\ell_0$ norm exploits sparsity directly, however, it is nonconvex and the problem in (\ref{formu:rankmin_CCS}) cannot be solved globally in practice. To avoid the nonconvexity and at the same time exploit sparsity, we utilize convex relaxation to relax the atomic $\ell_0$ norm. In particular, it can be relaxed in two ways from two different perspectives. One is to relax the atomic $\ell_0$ norm to the atomic $\ell_1$ norm (or simply the atomic norm) which is defined as the gauge function of $\text{conv}\sbra{\cA}$, the convex hull of $\cA$ \cite{chandrasekaran2012convex}:
\equ{\begin{split}\atomn{\m{Y}}
&\triangleq\inf\lbra{t>0: \m{Y}\in t\text{conv}\sbra{\cA}} \\
&= \inf\lbra{\sum_k c_k: \m{Y}=\sum_k c_k\m{a}_k, c_k\geq0, \m{a}_k\in\cA}.\end{split} \label{formu:atomicnorm}}
The atomic norm $\atomn{\cdot}$ is indeed a norm and convex by the property of the gauge function. The other way of convex relaxation is based on a perspective of rank minimization illustrated in (\ref{formu:AL0_rankmin}) and to relax the pseudo rank norm to the nuclear norm or equivalently the trace norm for a positive semidefinite matrix, i.e., to replace $\rank\sbra{\m{U}}$ by $\tr\sbra{\m{U}}$ in (\ref{formu:AL0_rankmin}). Interestingly enough, the two convex relaxations are equivalent, which is shown in the following result.

\begin{thm} $\norm{\m{Y}}_{\cA}$ defined in (\ref{formu:atomicnorm}) equals the optimal value of the following semidefinite programming (SDP):
\equ{\min_{\m{W},\m{u},\m{U}\geq\m{0}} \frac{1}{2\sqrt{N}}\tr\sbra{\m{U}}, \st \m{U}=\begin{bmatrix}\m{W} & \m{Y}^H \\ \m{Y} & T\sbra{\m{u}}\end{bmatrix}. \label{formu:AN_SDP}} \label{thm:AN_SDP}
\end{thm}

Theorem \ref{thm:AN_SDP} generalizes the results in \cite{candes2013towards,tang2012compressed} on the SMV case. Consequently, we propose the following atomic norm minimization problem for signal recovery:
\equ{\begin{split}
&\min_{\m{Y},\m{W},\m{u},\m{U}\geq\m{0}} \tr\sbra{\m{U}},\\
&\st \m{U}=\begin{bmatrix}\m{W} & \m{Y}^H \\ \m{Y} & T\sbra{\m{u}}\end{bmatrix}, \m{Y}_{\m{\Omega}}=\m{Y}^o_{\m{\Omega}}.\end{split} \label{formu:tracemin_CCS}}
It is shown in \cite{tang2012compressed} that $\m{Y}^o$ can be exactly recovered in the SMV case with high probability if $M\geq O\sbra{K\log K\log N}$ and the frequencies are separate by at least $\frac{4}{N}$. Note that the condition of frequency separation is introduced by the convex relaxation while it is not required in the atomic $\ell_0$ minimization problem as shown in Theorem \ref{thm:AL0_guanrantee}. After submission of this paper, we have proven in \cite{yang2014exact} that $\m{Y}^o$ can be recovered by (\ref{formu:tracemin_CCS}) in the MMV case under similar conditions.

\subsection{Iterative Reweighted Optimization via Nonconvex Relaxation} \label{sec:nonconvexrelax}

From the perspective of rank minimization, an iterative reweighted trace norm minimization scheme can be implemented to further improve the low-rankness by iteratively minimizing the objective function $\tr\mbra{\sbra{\m{U}_{j-1}+\epsilon\m{I}}^{-1}\m{U}}$ subject to the same constraints, where $\m{U}_{j}$ denotes the solution at the $j$th iteration starting with $\m{U}_0=\m{I}$, and $\epsilon>0$ is a small number. Obviously, the first iteration refers exactly to the convex relaxation. This iterative reweighted optimization scheme corresponds to relaxing $\rank\sbra{\m{U}}$ to the nonconvex objective $\ln\abs{\m{U}+\epsilon\m{I}}$, followed by a majorization-maximization (MM) implementation of the nonconvex optimization which guarantees local convergence of the objective function. We expect that a weaker condition of frequency separation holds for this nonconvex relaxation compared to the convex one since intuitively $\ln\abs{\cdot}$ is a closer approximation of the rank function. In the future, we can also consider other relaxation methods based on the literature of LRMC.

\subsection{Frequency and Amplitude Retrieval}
Given the solution of $\m{Y}$, it is of great importance to retrieve the frequency and amplitude solutions, or equivalently to obtain the atomic decomposition as in (\ref{formu:atomdecomp}), in applications such as line spectral estimation or DOA estimation. In particular, we can firstly obtain the frequency solution by the Vandermonde decomposition of $T\sbra{\m{u}}$ given the solution of $\m{u}$ (see details in \cite{yang2014gridless,yang2014exact}). Then the amplitude can be easily obtained by solving the linear system of equations in (\ref{formu:atomdecomp}).

\subsection{The Noisy Case} \label{sec:noisycase}
Noise is always present in practical scenarios. In this paper we consider only noise with bounded energy. Suppose the noise in the measurements $\m{Y}^o_{\m{\Omega}}$ is upper bounded by $\eta>0$ in the Frobenius norm. Then we can impose the inequality constraint $\frobn{\m{Y}_{\m{\Omega}}-\m{Y}^o_{\m{\Omega}}}\leq\eta$ instead of the equality constraint $\m{Y}_{\m{\Omega}}=\m{Y}^o_{\m{\Omega}}$ in the noiseless case. Note that the latter is a special case with $\eta=0$.

\subsection{Computationally Efficient Algorithms via ADMM}
We present a first-order algorithm based on ADMM to solve the trace minimization problems $\min\tr\sbra{\m{B}\m{U}}$ in Subsections \ref{sec:convexrelax} and \ref{sec:nonconvexrelax}, in particular,
\equ{\begin{split}
&\min_{\m{Y},\m{W},\m{u},\m{U}\geq\m{0}}\tr\sbra{\m{B}_{1}\m{W}} +\tr\sbra{\m{B}_{3}T\sbra{\m{u}}} +\tr\sbra{\m{B}_{2}^H\m{Y}+\m{Y}^H\m{B}_{2}},\\
&\st \m{U}=\begin{bmatrix}\m{W} & \m{Y}^H \\ \m{Y} & T\sbra{\m{u}}\end{bmatrix} \text{ and } \frobn{\m{Y}_{\m{\Omega}}-\m{Y}^o_{\m{\Omega}}}\leq\eta,\end{split} \label{formu:trmin_ADMM}}
where $\m{B}\triangleq\begin{bmatrix}\m{B}_1 &\m{B}_2^H \\\m{B}_2& \m{B}_3\end{bmatrix}\geq\m{0}$ is partitioned as $\m{U}$. (\ref{formu:trmin_ADMM}) can be solved within the framework of ADMM in \cite{boyd2011distributed}, where $\sbra{\m{Y},\m{W},\m{u}}$, $\m{U}$ and the Lagrangian multiplier are iteratively updated with closed-form expressions and converge to the optimal solution (see, e.g., \cite{yang2014gridless}). An eigen-decomposition of a Hermitian matrix of order $N+L$ is required at each iteration. We omit the detailed update rules due to the page limit.
Note that the ADMM converges slowly to an extremely accurate solution while moderate accuracy is typically sufficient in practical applications \cite{boyd2011distributed}.

\section{Numerical Simulations}
\begin{figure}
  \centering
  \includegraphics[width=3.4in]{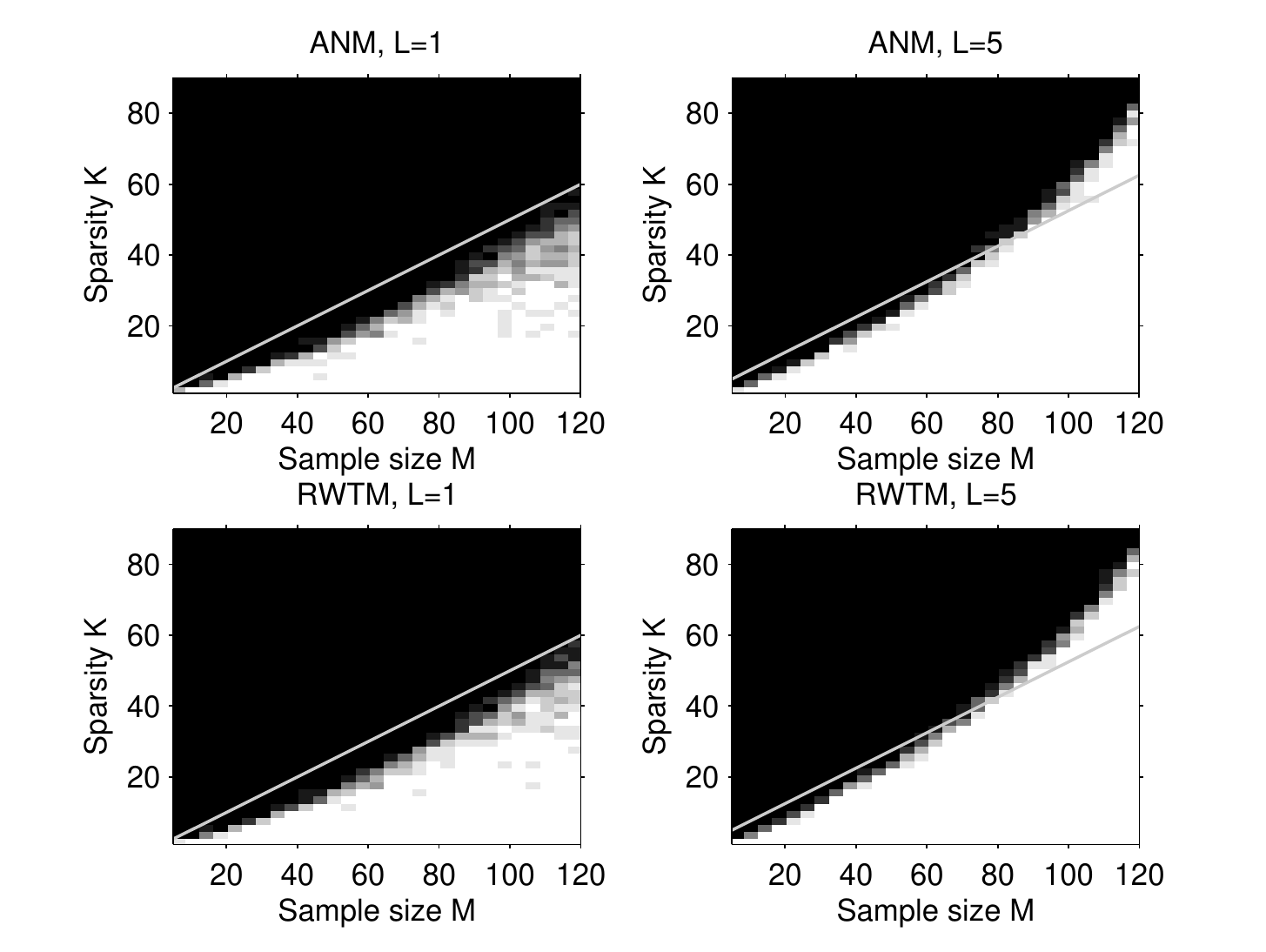}
  \caption{Phase transition with minimal frequency separation $\Delta_f\geq\frac{1}{N}$. White means complete success while black means complete failure. The straight lines correspond to $K=\frac{1}{2}\sbra{M+L}$.}
  \label{Fig:phasetransition}
\end{figure}

We first consider the noiseless case and study the so-called phase transition phenomenon in the $\sbra{M,K}$ plane. In particular, we repeat an experiment in \cite{tang2012compressed} and consider our proposed atomic norm (or trace norm) minimization (ANM) and reweighted trace minimization (RWTM) methods. RWTM is terminated within maximally 3 iterations in our simulation. For achieving high accuracy we solve the SDPs (in fact, their dual problems, see \cite{yang2014exact}) using a standard SDP solver, SDPT3 \cite{toh1999sdpt3}. We fix $N=128$ and vary $M=8,12,\dots,120$ and $K=2,4,\dots,M$. We consider $L=1$ and $5$, where ANM in the SMV case has been studied in \cite{tang2012compressed}. The frequencies $f_k$ are generated randomly with minimal separation $\Delta_f\geq1/N$ which is empirically found in \cite{tang2012compressed} to be the minimal separation required for exact recovery when $L=1$. The amplitudes $s_k\sbra{t}$ are randomly generated as $0.5+w^2$ with random phases, where $w$ is standard normal distributed. The first column of $\m{Y}^o$ is used in the SMV case in each problem generated. The recovery is considered successful if $\frobn{\widehat{\m{Y}}-\m{Y}^o}/\frobn{\m{Y}^o}< 10^{-6}$, where $\widehat{\m{Y}}$ denotes the recovered signal.\footnote{After submission of this paper, we find that this criterion is not strict enough to guarantee exact recovery of the frequencies. See more simulation results in \cite{yang2014exact}.} Simulation results are presented in Fig. \ref{Fig:phasetransition}, where a transition from perfect recovery to complete failure can be observed in each subfigure. By increasing the number of measurement vectors from 1 to 5, the phase of successful recovery is enlarged significantly for both ANM and RWTM. Moreover, RWTM has an enlarged success phase than ANM, especially in the MMV case, due to adoption of nonconvex relaxation. We also notice that the transition boundary of ANM with $L=1$ is not very sharp and failures happen in the area where complete success is expected. Further examination reveals that most of the failures happen when the minimal separation marginally exceeds $1/N$. The situation is better for RWTM with $L=1$. In contrast, sharp phase transitions exhibit for both ANM and RWTM in the MMV case. This implies that the requirement of frequency separation can be relaxed in our considered MMV case where the measurement vectors are statistically independent.

We also plot the line $K=\frac{1}{2}\sbra{M+L}$ in each subfigure which acts as an upper bound of the sufficient condition in Theorem \ref{thm:AL0_guanrantee} for exact recovery using the atomic $\ell_0$ norm minimization. We see that in the MMV case successful recoveries can be obtained even above the line with ANM or RWTM, implying that the sufficient condition is unnecessary.


\begin{figure}
  \centering
  \includegraphics[width=3in]{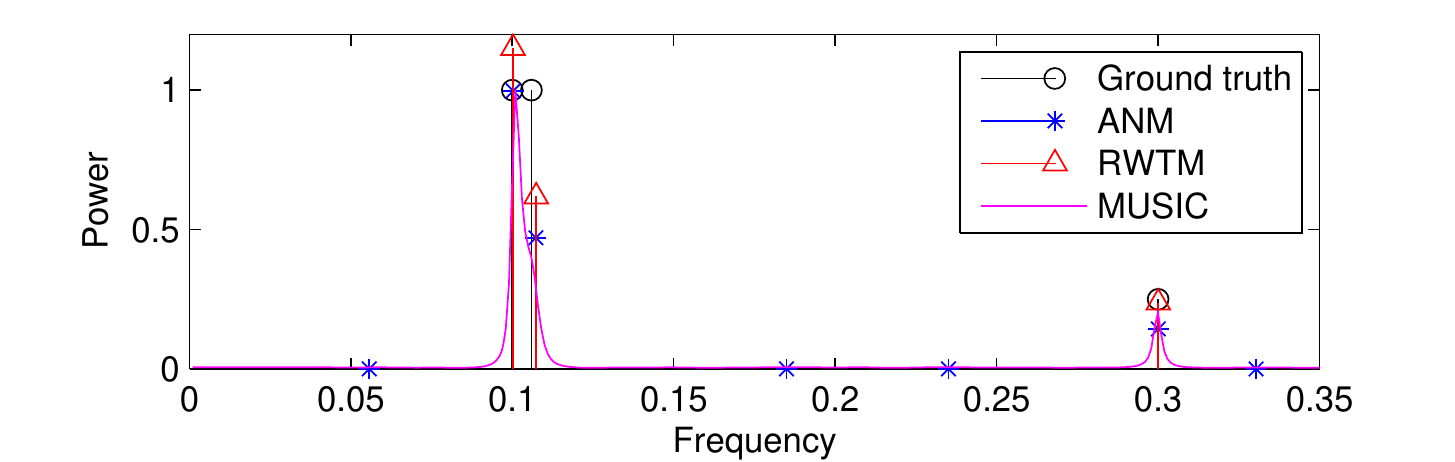}
  \caption{DOA estimation using ANM and RWTM compared to MUSIC (shown only on the frequency interval $[0,0.35]$).}
  \label{Fig:comparewithMUSIC}
\end{figure}

We next consider an example of DOA estimation using a redundancy sparse linear array (SLA) of size $M=10$ with $\m{\Omega}=[1,2,  7,11, 24, 27,    35,    42,    54,    56]^T$ (see, e.g., \cite{yang2014discretization}). By assuming narrowband sources the DOA estimation problem is mathematically equivalent to frequency recovery in continuous CS. We apply the proposed ANM and RWTM methods with the ADMM implementations to the DOA estimation and compare with MUSIC. In the simulation, we consider $K=3$ sources impinging on the array from directions corresponding to frequencies $\m{f}=[0.1, 0.106, 0.3]^T$ with powers 1, 1, and 0.25. Suppose that $L=10$ snapshots (or measurement vectors) are observed with the signal to noise ratio $\text{SNR}=14.2$dB. The DOA estimation results of ANM and RWTM are presented in Fig. \ref{Fig:comparewithMUSIC} compared to MUSIC. It is shown that both ANM and RWTM can separate the first two sources while MUSIC cannot. Note also that ANM produces a few spurious sources with very small powers while RWTM detects exactly 3 sources. Both ANM and RWTM take about 2 seconds (loose convergence criteria are adopted in the first few iterations of RWTM for speed acceleration). Finally, it is worth noting that ANM and RWTM require the knowledge of the noise level while MUSIC needs to know the source number.

\section{Conclusion}
In this paper, the SMV and MMV continuous CS problems were studied in a unified framework and linked to low rank matrix completion. We extended existing discrete CS results to the continuous case, introduced computationally efficient algorithms and validated their performances via simulations. We have recently analyzed the proposed atomic norm minimization method in \cite{yang2014exact}, which generalizes \cite{candes2013towards} and \cite{tang2012compressed}.



\end{document}